\documentclass[conference,letterpaper]{IEEEtran}
\IEEEoverridecommandlockouts

\usepackage[utf8]{inputenc} 
\usepackage[T1]{fontenc}
\usepackage{url}
\usepackage{ifthen}
\usepackage{cite}
\usepackage[cmex10]{amsmath}
\usepackage{algorithmic}
\usepackage{amssymb}
\usepackage{algorithm}
\usepackage{xcolor}
\usepackage{mathtools}
\usepackage{soul}
\usepackage[normalem]{ulem}

\usepackage[caption=false,font=footnotesize]{subfig}

\usepackage{thmtools, thm-restate}

\newcounter{remarkcnt}
\newenvironment{boldremark}[1][]{\refstepcounter{remarkcnt}\par\smallskip
   \noindent \textbf{Remark~\theremarkcnt. #1} \rmfamily}{\smallskip}

\newcommand{\vh}{\boldsymbol{h}}
\newcommand{\vk}{\boldsymbol{k}}
\newcommand{\hyp}{\boldsymbol{h}^t_{\boldsymbol{k}}}

\newcommand{\vz}{\boldsymbol{z}}

\begin{document}
\title{Enhancing Binary Search via Overlapping Partitions}

\author{\IEEEauthorblockN{Kaan Buyukkalayci$^\dagger$, Merve Karakas$^\dagger$, Xinlin Li$^\dagger$, and Christina Fragouli$^\dagger$\\ 
$^\dagger$University of California, Los Angeles\\
Email:\{kaanbkalayci, mervekarakas, xinlinli, christina.fragouli\}@ucla.edu}\thanks{The work was supported in part by NSF award 2146828 and the Army Research Laboratory grant under Cooperative Agreement W911NF-17-2-0196.}}

\maketitle

\begin{abstract}
This paper considers the task of performing binary search under noisy decisions, focusing on the application of target area localization. In the presence of noise, the classical partitioning approach of binary search is prone to error propagation due to the use of strictly disjoint splits. While existing works on noisy binary search propose techniques such as query repetition or probabilistic updates to mitigate errors, they often lack explicit mechanisms to manage the trade-off between error probability and search complexity, with some providing only asymptotic guarantees. To address this gap, we propose a binary search framework with tunable overlapping partitions, which introduces controlled redundancy into the search process to enhance robustness against noise. We analyze the performance of the proposed algorithm in both discrete and continuous domains for the problem of area localization, quantifying how the overlap parameter impacts the trade-off between search tree depth and error probability. Unlike previous methods, this approach allows for direct control over the balance between reliability and efficiency. Our results emphasize the versatility and effectiveness of the proposed method, providing a principled extension to existing noisy search paradigms and enabling new insights into the interplay between partitioning strategies and measurement reliability. 
\end{abstract}

\section{Introduction}
Binary search is a classical algorithm that iteratively partitions a population of size $n$ to efficiently identify a target element.
While its simplicity and logarithmic complexity make it highly popular, it may not perform well in applications where the decision criteria used to guide the search are corrupted by noise. This limitation arises in a wide range of real-world scenarios, including but not limited to medical diagnostics, sensor readings, recommendation systems, and crowdsourcing tasks where human responses may be inconsistent or erroneous.

\begin{figure}[t]
    \centering
    \subfloat[Non-overlapping partitions.\label{fig:non_overlap_tree}]{
        \includegraphics[width=0.45\columnwidth]{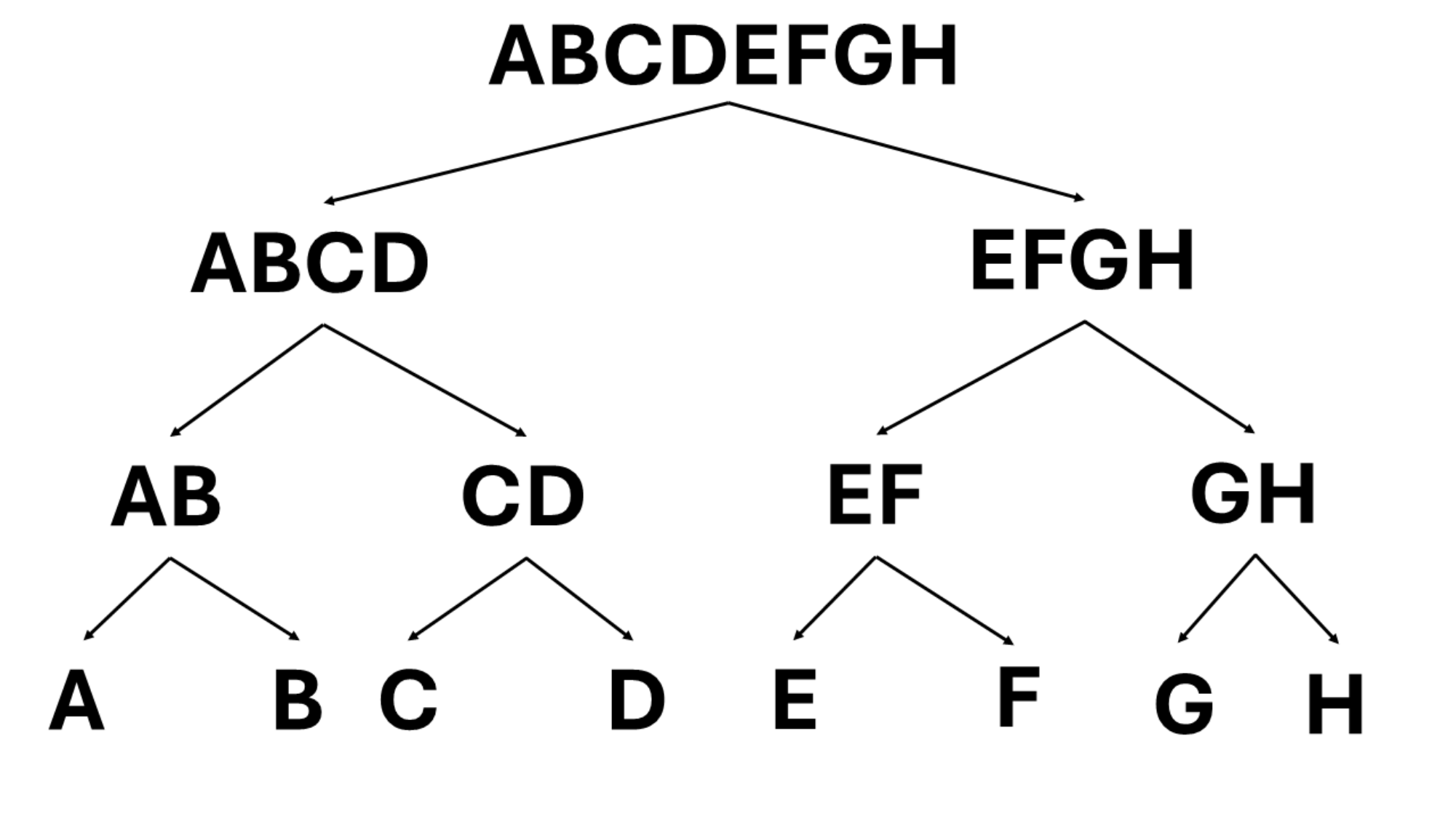}
    }
    \hfill
    \subfloat[Overlapping partitions.\label{fig:overlap_tree}]{
        \includegraphics[width=0.45\columnwidth]{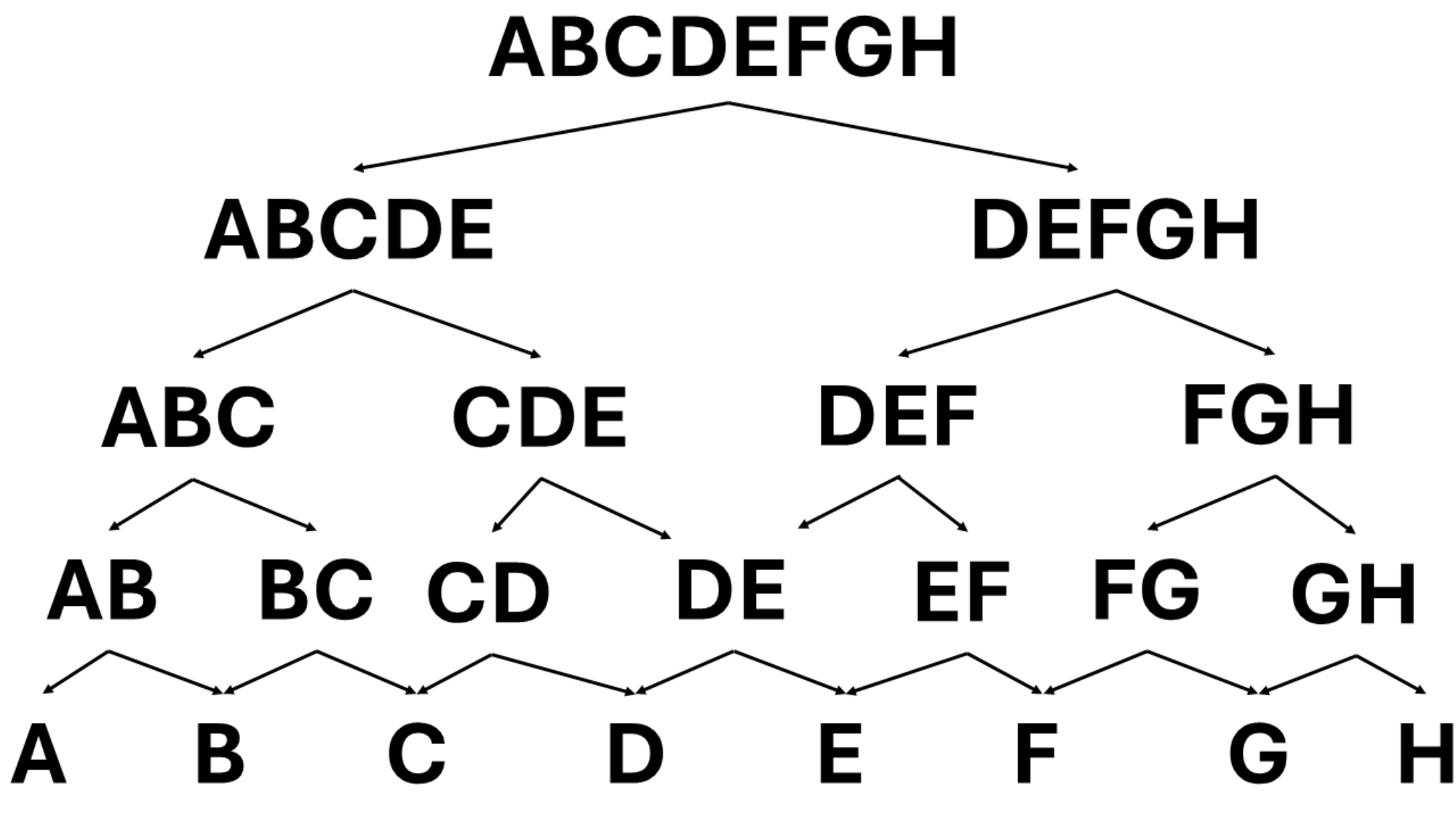}
    }

    \caption{Examples of decision trees for $n=8$ elements.}
    \label{fig:tree_comparison}
\end{figure}

To address this limitation, we explore a \emph{modified binary search algorithm} that incorporates \emph{overlapping regions} between the examined parts at each step. By introducing overlap, the algorithm ensures that the target element is less likely to be excluded due to noise, thereby increasing robustness at the cost of additional search steps.
%\textcolor{blue}{Here we need a figure with a tree say with n=16 and no overlap, then the tree of how it becomes with overlap. And a small math analysis of the tradeoff, potentially a figure with a toy example. Add a  paragraph and a trade-off figure along the lines:
%(Calculation that illustrates overlap as a percentage wihth figure of binary tree Pet = 0.5 / (something for degree or overlap or alpha) what is the best tradeoff depending on the application and we are going to do this for this particular application)
% }
We illustrate this idea through a toy example. Consider a binary search tree with $n$ leaves as in the example depicted in Fig. \ref{fig:non_overlap_tree} for $n=8$. With no overlap, the tree depth equals $\log_2 (n)=3 $.
%and each node (partition of elements) at depth $t$ from the root contains $n^{(t)}=n^{(t-1)}/2=n/2^t$ elements. 
Assume now that we partition with some overlap as illustrated in Fig.~\ref{fig:overlap_tree}
with the result that the depth of the tree now increases to $4$.

%that an overlap parameter \( \alpha \) controls the  redundancy in the partitions,  namely $n^{(t)}=\operatorname{ceil}\left(\left(\frac{1}{2}+\alpha\right)n^{(t-1)}\right)$ \textcolor{blue}{Kaan:I dont think ceil() works for our example in Fig 2. Then we cant go from 5 to 3 and 2 to 1} where $\operatorname{ceil(\cdot)}$ is the ceiling operator that rounds it argument to the next largest integer. 
 %The depth of the tree is now $\beta(\alpha) = \gamma(\alpha)\log_2(n)$ where $\gamma(\alpha) \geq 1$ is a constant that does not depend on $n$ and can be lower bounded as  ${1}/{\log(\frac{1}{2}+\alpha)}$. %\textcolor{blue}{Please check this.Kaan: It woudl have worked if we could use ceil() but we can just say approximately?} For instance, in Fig.~\ref{fig:overlap_tree} we use $\alpha=1/8$, which results in a tree of depth 4.

This alternate structure can be attractive in  applications where the probability of decision error 
improves by allowing overlap.  In some applications, we may also not be interested in uniquely identifying the target element, but instead, simply wish to narrow down its location within a set of candidates, akin to list decoding. In such cases, we can use the following parameters to describe the achieved tradeoffs: $P_e$, the probability of error, \( \beta \), the number of decision steps we take (depth of the tree we traverse) and $|\mathcal{H}^{\beta}|$, the size of the set of potential target elements at the algorithm's output.

In this paper we explore such an application, motivated from the practical problem of localizing a target in a two-dimensional search space. To do so, we want to use noisy measurements from sensors deployed in a two-dimensional field to take overlapping binary search decisions. In particular, at each step, the search space is divided into two overlapping halves along its longest dimension, and Maximum Likelihood (ML) estimation is used to select the half most likely to contain the target. 
Our main contributions are as follows: 
\begin{enumerate}
    \item We introduce a binary search framework with overlapping partitions of the search space  and explore its application to target area localization. 
    \item We analytically derive the error probability for both discrete and continuous search spaces by leveraging Voronoi regions in the sensor measurement space.
    \item  We explore the trade-off between the amount of overlap, probability of error, and delay in the above scenario through numerical evaluation.
\end{enumerate}
% \textcolor{to w}

\noindent\textbf{Related Work.} Dealing with noise in classical binary search dates back to Horstein's probabilistic bisection procedure in the context of channel coding with feedback \cite{Sequentialtransmissionusingnoiselessfeedback} which was shown to be asymptotically optimal \cite{Anintervalestimationproblemforcontrolledobservations} \cite{Noisybinarysearchanditsapplications} in terms of minimizing the expected search tree depth by guaranteeing optimal decay of error probability under a noisy binary search model for a single-dimensional quantity. This approach is later generalized to more general search spaces in \cite{NoisyGeneralizedBinarySearch}. However, the proposed mechanisms do not allow managing the tree depth-error probability tradeoff in finite-horizon settings.
 Another line of work \cite{NearOptimalBayesianActiveLearningforDecisionMaking} explores overlapping decision regions within a hypothesis space, where the goal is to converge to a predefined decision region as quickly as possible; yet unlike our work their decision regions are fixed. Several well-established studies on binary search address noise by employing repeated queries \cite{Noisybinarysearchanditsapplications}\cite{ADirectSearchAlgorithmforOptimizationwithNoisyFunctionEvaluations} \cite{TheGeometryofGeneralizedBinarySearch} to disjoint partitions of the search space.

For the localization application we consider, much of the literature focuses on applying signal processing techniques—often rooted in maximizing Fisher information—to minimize mean squared error (MSE). Representative works (see \cite{fzwwf2018, n2021, ylw2013,mb2006, bfadp2010, tkh2007, dqcs2021, slv2014, flt2012, bbsjr2015, h2018}) analyze optimal sensor-target geometries, multi-target tracking formulations in large-scale problems, or Bayesian estimation under Gaussian priors. Although these studies cover a wide range of geometric designs and adaptive scheduling schemes, they mostly look at exact localization (not within an area) and do not leverage binary search mechanisms.

\noindent \textbf{Paper Organization.} The rest of the paper is structured as follows. Section~\ref{sec:problemformulation} introduces the notation, problem formulation, and the iterative search framework. Section~\ref{sec:approach} describes our overlapping binary search algorithm. Section~\ref{sec:analysis} provides an analysis of the error probability over both discrete and continuous search spaces. Section ~\ref{sec:evaluation} provides numerical results. Finally, Section~\ref{sec:conclusion} concludes the paper.

\section{Problem Formulation}
\label{sec:problemformulation}
%\textcolor{blue}{Not sure if we need a figure here}
%\mk{alternative problem formulation}
We consider an overlapping binary search framework to locate a stationary signal source within a region $\mathcal{H}^0 \subseteq \mathbb{R}^d$.  Let $\boldsymbol{l}_T \in \mathcal{H}^0$ denote the true (but unknown) location of the target.  A large number of sensing assets (or sensors) are distributed across $\mathcal{H}^0$, from which we can selectively \emph{activate} a subset at each step~$t$ to reduce our uncertainty about $\boldsymbol{l}_T$.

\noindent\textbf{Measurement Model.}
Let $\mathcal{S}^t = \{\boldsymbol{s}_1^t,\dots,\boldsymbol{s}_{|\mathcal{S}^t|}^t\}$ be the locations of activated sensors at step~$t$, where $|\mathcal{S}^t|$ may be constrained by resource limits.  Each sensor $\boldsymbol{s}_j^t$ measures the signal:
\begin{equation}
%\label{eq:measurement}
    z_j^{t} \;=\; f\!\Bigl(\|\boldsymbol{s}_j^t - \boldsymbol{l}_T\|_2\Bigr) \;+\; \eta_{j}^t, \nonumber
\end{equation} where $f(\cdot)$ models the expected measurement as a function of distance to the target, and $\eta_{j}^t \sim \mathcal{N}(0,\sigma^2)$ is additive white Gaussian noise. When $z_j^t$ represents, for example, received signal strength (measured in dB) and $f(\cdot)$ models $\log$-distance decay, this framework also captures scenarios with log-normal uncorrelated noise, common in outdoor wireless localization.  In vector form, we write $\boldsymbol{z}^t = \bigl[z_1^t,\dots,z_{|\mathcal{S}^t|}^t\bigr]^\top$.
For each hypothesis (i.e., possible location of the target) $\vk$,
we use $\vh_{\boldsymbol{k}}^t  \in \mathbb{R}^{|\mathcal{S}^t|}$ to denote the expected measurements from all activated sensors, where $h_{\boldsymbol{k},i}^t = f\left(\|\boldsymbol{s}_i^t - \boldsymbol{k}\|_2\right)$ is the $i$-th element of vector $\vh^t_{\vk}$.

%$\vh_{\boldsymbol{k}}^t$ denote the \emph{noise-free} measurement vector for hypothesis $\boldsymbol{k}$ (i.e., a potential target location).}

\noindent\textbf{Iterative Search.}
At each step~$t$, the algorithm maintains a \emph{candidate region} (or “search space”) $\mathcal{H}^{t-1}$ containing all possible target locations still under consideration.  By activating sensors $\mathcal{S}^t$ and collecting $\boldsymbol{z}^t$, we refine our knowledge of $\boldsymbol{l}_T$ and \emph{shrink} the search region to $\mathcal{H}^t \subseteq \mathcal{H}^{t-1}$.  This continues for $\beta$ steps or until the region is sufficiently small.

\noindent\textbf{2D Digitized Search Space.}
For concreteness and to keep notation manageable, we focus on a two-dimensional domain subdivided into a grid and present our analysis and proposed algorithms under this setting, though our methods extend directly to general $d$-dimensional (continuous) spaces.  Concretely, assuming a rectangle search space in $\mathbb{R}^2$, we use $(L_x^t, L_y^t)$ to denote the side length of $\mathcal{H}^t$ and $(n_x^{(t)},n_y^{(t)})$ are the grid resolutions in each dimension. The initial search space is given as
%\small
\begin{align}
\mathcal{H}^0 = &\Bigr\{ \vk = \left( \frac{(2i-1)L_x^0}{2n_x^{(0)}}, \frac{(2j-1)L_y^0}{2n_y^{(0)}} \right) \;\Bigr|\; i\in [n_x^{(0)}],j \in [n_y^{(0)}] \Bigr\} \nonumber
\label{eq:Search Space}
\end{align}
%\normalsize
Each point in $\mathcal{H}^0$ is a potential location for $\boldsymbol{l}_T$, and the selection of $\{{n^{(0)}_{x}},n_y^{(0)}\}$ controls the coarseness of the grid.

\noindent\textbf{Performance Criteria.}
We explore the trade-off among three main objectives:
\begin{enumerate}%[leftmargin=2em]
    \item \textit{Search depth} $(\beta)$: the total number of steps taken before stopping;
    \item \textit{Final region size} $(|\mathcal{H}^\beta|)$: the cardinality or volume of the search space after $\beta$ steps;
    \item \textit{Error probability} $(P_e)$: the probability that the true target location is \emph{excluded} from $\mathcal{H}^\beta$.  Specifically,
    \begin{equation}
    \label{eq:tot_error}
        P_e \;=\; \Pr\!\bigl\{\boldsymbol{l}_T \notin \mathcal{H}^\beta\bigr\}
        \;=\; 1 \;-\;\prod_{t=1}^{\beta}\bigl(1 - P_e^t\bigr),
    \end{equation}
    where $P_e^t = \Pr\{\boldsymbol{l}_T \notin \mathcal{H}^t \mid \boldsymbol{l}_T \in \mathcal{H}^{t-1}\}$ is the step-$t$ misclassification probability.
\end{enumerate}

\section{Binary Search with Overlapping Partitions}
\label{sec:approach} 
This section formally describes an iterative binary search procedure
that, starting from an initial region $\mathcal{H}^0$, reduces it at each step $t$ to a candidate region $\mathcal{H}^t \subseteq \mathcal{H}^{t-1}$ and stops after $\beta$
such steps  yielding a refined candidate region \( \mathcal{H}^\beta \)
(see pseudocode in Algorithm \ref{alg:iterative_binary_search}).
At each step $t$, Algorithm 1 divides $\mathcal{H}^{t-1}$  into two overlapping 
candidate regions  $\{\mathcal{H}^{t}_1, \mathcal{H}^{t}_2\}$, where the overlap is controlled by a parameter \( \alpha \), and selects one of them.
In particular, let the longest dimension of the  search space $\mathcal{H}^{t-1}$ be 
\begin{equation}
    d^t = \arg \max_{i\in\{1,2\}} \left \{\max_{\vk\in\mathcal{H}^{t-1}} k_i - \min_{\vk\in\mathcal{H}^{t-1}} k_i \right \} \nonumber
\end{equation}
and calculate the midpoint along $d^t$ as
\begin{equation}
m_{\mathrm{d}^t}^t = \frac{1}{2}\left(\max_{\vk\in\mathcal{H}^{t-1}} k_{\mathrm{d}^t} + \min_{\vk\in\mathcal{H}^{t-1}} k_{\mathrm{d}^t}\right). \nonumber
\end{equation}
Algorithm 1 considers the candidate regions

\begin{equation}\label{eq:subregion}
\begin{aligned}
\mathcal{H}_1^t = & \{\boldsymbol{k} \in \mathcal{H}^{t-1} \mid k_{\mathrm{d}^t} \leq m_{\mathrm{d}^t}^t + \delta^t \Delta_{\mathrm{d}^t}\}, \\
\mathcal{H}_2^t = & \{\boldsymbol{k} \in \mathcal{H}^{t-1} \mid k_{\mathrm{d}^t} \geq m_{\mathrm{d}^t}^t - \delta^t \Delta_{\mathrm{d}^t}\},
\end{aligned}
\end{equation}
% \begin{equation}
% \mathcal{H}_1^t = \{\boldsymbol{k}_i \in \mathcal{H}^t \mid k_{i,\mathrm{d}^t} \leq m_{\mathrm{d}^t}^t + \delta^t \Delta_{\mathrm{d}^t}\},
% \end{equation}
% \begin{equation}
% \mathcal{H}_2^t = \{\boldsymbol{k}_i \in \mathcal{H}^t \mid k_{i,\mathrm{d}^t} \geq m_{\mathrm{d}^t}^t - \delta^t \Delta_{\mathrm{d}^t}\},
% \end{equation}
where \( \Delta_{\mathrm{d}^t} = \frac{L_{\mathrm{d}^t}^{0}}{n^{(0)}_{{\mathrm{d}^t}}} \) represents the grid step size along the splitting direction . \( \delta^t \) is computed so that

\small
\begin{equation}
n_{d^t}^{(t)} = \max \left(\operatorname{round}\left(\left(\frac{1}{2} + \alpha\right)n^{(t-1)}_{d^t}\right),\operatorname{ceil}\left(\frac{1}{2} n^{(t-1)}_{d^t}\right)\right), \label{eq:delta_identifier}
\end{equation} \normalsize
where \( \alpha \in \left[0,\frac{1}{4}\right)\) controls the degree of overlap, \( n_{d^t}^{(t)} \) denotes the number of grid points in the splitting direction at step \( t \). \( \operatorname{round}(\cdot) \), which rounds  to the nearest integer with ties resolved  to the nearest even integer, and $\operatorname{ceil(\cdot)}$ functions here are used to tackle special cases in integer-related issues; the restriction $\alpha < 1/4$ ensures convergence for small $n_{d^t}^{(t)}$.
\begin{algorithm}[t!]
\caption{Overlapping Binary Search for Area Localization}
\label{alg:iterative_binary_search}
\begin{algorithmic}[1]
{
\STATE \textbf{Inputs:} Search Space $\mathcal{H}^0$, overlapping degree $\alpha$, number of search steps $\beta$, signal propagation function $f(\cdot)$
\FOR{\( t \leftarrow 1, \dots, \beta \)}
    \STATE Partition $\mathcal{H}^{t-1}$ into overlapping sub-regions $\{\mathcal{H}^{t}_1, \mathcal{H}^{t}_2\}$ as defined in \eqref{eq:subregion}
    % \STATE \( \mathrm{d}^t = \arg\max_{d \in \{x, y\}} L_d^t,\)
    % \STATE \( m_{\mathrm{d}^t}^t = \frac{1}{2} \left( \min_i k_{i,\mathrm{d}^t} + \max_i k_{i,\mathrm{d}^t}\right) \)
    % \STATE \( \mathcal{H}_1^t = \{\boldsymbol{k}_i \in \mathcal{H}^{t-1} \mid k_{i,\mathrm{d}^t} \leq m_{\mathrm{d}^t}^t + \delta^t \Delta_{d^t}\} \)
    % \STATE \( \mathcal{H}_2^t = \{\boldsymbol{k}_i \in \mathcal{H}^{t-1} \mid k_{i,\mathrm{d}^t} \geq m_{\mathrm{d}^t}^t - \delta^t
    % \Delta_{d^t}\} \)
    \STATE Activate $|\mathcal{S}^t|$ sensors and obtain measurements $\vz^t$
    %\STATE Obtain $|\mathcal{S}^t|$ measurements $z_j^t, ~ j = 1, \dots |S^t|$
    \STATE Update the likelihood $\ell(\boldsymbol{k}), \forall \vk \in \mathcal{H}^{t-1}$ as in \eqref{eq:mle}
    % \STATE \( \ell(\boldsymbol{k}) = \sum_{j=1}^{|\mathcal{S}^t|} \log p(z_j^t \mid f(\boldsymbol{k}_i, \boldsymbol{s}_j), \sigma^2), \quad \forall \boldsymbol{k}_i \in \mathcal{H}^{t-1} \)
    % \STATE \( \boldsymbol{k}_{\mathrm{ML}}^t = \arg\max_{\boldsymbol{k}_i \in \mathcal{H}^t} \ell(\boldsymbol{k}_i) \)
    % \STATE \( \mathcal{H}^{t+1} = 
    % \begin{cases} 
    %     \mathcal{H}_1^t & \text{if } \boldsymbol{k}_{\mathrm{ML}}^t \in \{\boldsymbol{k}_i \in \mathcal{H}^t \mid k_{i,\mathrm{d}^t} \leq m_{\mathrm{d}^t}^t\}, \\
    %     \mathcal{H}_2^t & \text{otherwise}
    \STATE $\mathcal{H}^{t} \leftarrow \arg \max_{i\in\{1,2\}} \left\{\max_{\vk \in \mathcal{H}^t_i} \ell(\vk)\right\}$
    % \STATE Apply decision rule \eqref{eq:rule2} to find $\mathcal{H}^t$.
\ENDFOR
\STATE \textbf{Output:} \( \mathcal{H}^\beta \)
}
\end{algorithmic}
\end{algorithm}
{ It then uses measurements $\vz^t$ from a selected set of sensors $\mathcal{S}^t$  to update the search space 
by retaining the most likely region to contain the target, based on Maximum Likelihood (ML) estimation. For each possible target location $\vk \in \mathcal{H}^{t-1}$, it calculates the likelihood $\ell(\vk)$ as

\begin{equation}\label{eq:mle}
     \ell(\vk) = \sum_{j=1}^{|\mathcal{S}^t|} \log p(z_j^t \mid \boldsymbol{l}_T=\vk), \quad \forall \boldsymbol{k} \in \mathcal{H}^{t-1},
\end{equation}
 }
and selects the region with
\begin{equation}\label{eq:rule2}
\mathcal{H}^{t} = \arg \max_{i\in\{1,2\}} \left\{\max_{\vk \in \mathcal{H}^t_i} \ell(\vk)\right\}.
\end{equation}

Note that other selection criteria, e.g., $\mathcal{H}^{t} = \arg \max_{i\in\{1,2\}} \sum_{\vk \in \mathcal{H}^t_i} \ell(\vk)$ can also be used.

\section{Main Results (Error Probability)}
\label{sec:analysis}
In this section we  characterize the error probability of Algorithm~\ref{alg:iterative_binary_search}.
We argue that 
%the error probability can be reduced to a problem in computational geometry.   show that
the error event at step~\(t\) can be analyzed by 
using Voronoi regions in the 
\(\lvert S^t \rvert\)-dimensional  sensor measurements space. 
%While some arguments that view hypothesis sets as regions in multidimensional space have previously been applied in generalized binary search techniques~\cite{NoisyGeneralizedBinarySearch, NearOptimalBayesianActiveLearningforDecisionMaking, TheGeometryofGeneralizedBinarySearch}, our setting here differs in that the "similarity" between different hypotheses $\vk$ is readily apparent by their proximity in 2-dimensional space, akin to classical binary search.

\subsection{Voronoi Description of the Error Event}
\begin{restatable}[Geometry of the Error Event]{lemma}{thmMain}\label{thm:main}
For the case where $\boldsymbol{l_T} \notin \mathcal{H}^t_1 \cap \mathcal{H}^t_2$, assume \(\mathcal{H}_m^t \subseteq \mathcal{H}^{t-1}\)  contains the true target location \(\boldsymbol{l}_T\), that is, \(m \in \{1, 2\}\) is the index of the correct hypothesis set.  Then the error probability at step \(t\) is given by
\begin{equation}
P_e^t \;=\; \Pr\!\biggl(\,\boldsymbol{z}^t \,\notin\, 
\bigcup_{\vk \in \mathcal{H}_m^t} V_{\vk}^t\biggr), \nonumber
\end{equation}
where \(V_{\vk}^t\) is the Voronoi cell corresponding to \(\vh_{\vk}^t\) among all 
\(\{\vh_{\vk}^t : \vk \in \mathcal{H}^{t-1}\}\), namely,
{\small
\begin{equation} \nonumber
V_{\vk}^t 
\;=\; 
\Bigl\{\,\boldsymbol{x}\in \mathbb{R}^{\lvert S^t\rvert}\,\Bigm|\,
\|\boldsymbol{x} - \vh_{\vk}^t\|_2 \,\le\, \|\boldsymbol{x} - \vh_{\vk'}^t\|_2,\; \forall \,\vk' \neq \vk \in \mathcal{H}^{t-1}
\Bigr\}.
\end{equation}
}
Note that if $\boldsymbol{l_T} \in \mathcal{H}^t_1 \cap \mathcal{H}^t_2$, selecting either of the hypothesis subsets is correct, which implies that $P_e^t = 0$.
\end{restatable}
\noindent \textit{Proof Sketch} (see Appendix~\ref{appx:thm-main}\cite{Appendix} for full proof):
%\noindent\textit{Proof Sketch (see Appendix~\ref{appx:thm-main} for details).}
Under independent Gaussian noise with equal variance, the likelihood function reduces to finding $\vk \in \mathcal{H}^{t-1}$ such that \(\|\boldsymbol{z}^t - \vh_{\vk}^t\|_2\) is minimized.  Hence, the ML decision rule assigns the observed \(\boldsymbol{z}^t\) to whichever \(\vh_{\vk}^t\) is nearest in Euclidean distance.  This induces a Voronoi partition 
\(\{V_{\vk}^t\}\).  If $\boldsymbol{l}_T \notin \mathcal{H}_1 \cap \mathcal{H}_2$, an error occurs if \(\boldsymbol{z}^t\) falls into a Voronoi cell corresponding to a \(\vk \notin \mathcal{H}_m^t\).  
The probability of that event is exactly 
\(\Pr(\boldsymbol{z}^t \notin \cup_{\vk \in \mathcal{H}_m^t} V_{\vk}^t)\).
\smallskip
\begin{boldremark}
To compute the error probability in Lemma~\ref{thm:main}, recall that the observation space is partitioned into Voronoi cells 
$\{V_{\vk}^t\}_{\vk \in \mathcal{H}^{t-1}}$, each defined by linear inequalities (since they can equivalently be expressed as unions of halfspaces) and hence representable as a convex polyhedron.  
The union 
$\bigcup_{\vk \in \mathcal{H}_m^t} V_{\vk}^t$
corresponds to the set of measurement vectors $\boldsymbol{z}^t$ that would lead to the correct decision at time $t$. 

Well-known computational-geometry algorithms (e.g., \cite{fortune1987,guibas1985,
aurenhammer1991, deberg2008}) can be used to determine or approximate the boundaries of Voronoi partitions in high-dimensional space, which subsequently allows us to compute the boundary of the union $\bigcup_{\vk \in \mathcal{H}_m^t} V_{\vk}^t$.
%\textcolor{blue}{ by locating neighboring pairs of points where one belongs to $\bigcup_{\vk \in \mathcal{H}_1^t} V_{\vk}^t$ and the other to $\bigcup_{\vk \in \mathcal{H}_2^t} V_{\vk}^t$. Kaan: Not sure if this is clear}
Once that region’s boundary is determined, we can integrate the Gaussian density of $\boldsymbol{z}^t$ over either the union itself (for the probability of a correct decision) or its complement (for $P_e^t$) for each hypothesis $\vk$. Although the noise is uncorrelated across sensors, one still faces a \emph{multivariate} Gaussian integral because the ensemble decision boundary is piecewise linear. In practice, this integral can be evaluated numerically or approximated via known techniques.

Nevertheless, such piecewise-linear boundaries can be unwieldy for direct analysis in higher dimensions. In the following subsection, we exploit a \emph{symmetric} sensor placement strategy to reduce these boundaries to \emph{simple hyperplanes}, thereby providing a closed-form expression for $P_e^t$ in this special but practically relevant scenario (see Thm.~\ref{thm:symm}).
\end{boldremark}

\subsection{A Simplified Formula under Symmetric Measurements}
\label{sec:symmetric}
A notable special case arises if the sensors we activate at step \(t\) are placed symmetrically around some midpoint in the “splitting dimension.”  This symmetry yields a simple closed form for the error probability, as shown next.
\begin{restatable}{theorem}{thmSym}\label{thm:symm}
Let \(\mathcal{S}^t = \{\boldsymbol{s}_1^t, \boldsymbol{s}_2^t\}\) be placed symmetrically around the midpoint \(m_{d^t}^t\) in dimension \(d^t\), and assume a uniform prior on \(\mathcal{H}^{t-1}\). The error probability at step \(t\) satisfies
\[
P_e^t 
\;=\; 
\frac{2}{n_x^{(t-1)}\,n_y^{(t-1)}} 
\sum_{\vk \,\in\, \mathcal{H}_1^t \setminus \mathcal{H}_2^t} 
Q\!\Bigl(\tfrac{d_{\vk}}{\sigma}\Bigr),
\]
where \(Q(\cdot)\) is the standard Gaussian tail function and $d_{\vk}$ is the distance of $\vk$ to the decision boundary.
\end{restatable}
 \noindent{\em Note:}  $d_{\vk}$ can be computed as the half distance from $\vh_{\vk}$ to $\vh_{\vk'}$, where $\vk'$ is the symmetric hypothesis point to $\vk$ in the dimension of $d^t$ (see detailed description in Appendix B  \cite{Appendix}).

\smallskip
\noindent \textit{Proof Sketch} (full proof is in Appendix~\ref{appx:thm-symm} \cite{Appendix}):
Symmetry implies that each hypothesis \(\vk \in \mathcal{H}_1^t\) has a "partner" \(\vk' \in \mathcal{H}_2^t\) such that the distance of \(\vk\) to \(\boldsymbol{s}_1^t\) equals the distance of \(\vk'\) to \(\boldsymbol{s}_2^t\), and vice versa.  Consequently, the ensemble decision boundary is a single hyperplane in \(\mathbb{R}^{\lvert S^t\rvert}\).  Projecting the noise onto the normal of this hyperplane reduces to a univariate Gaussian, yielding the closed-form sum of \(Q\)-functions.

\smallskip
The next corollary generalizes from a discrete set \(\mathcal{H}^{t-1}\) to a continuous region in \(\mathbb{R}^2\) by allowing \(|\mathcal{H}^{t-1}|\to\infty\) and replacing the finite sum by a Riemann integral.  

\smallskip
\begin{restatable}{corollary}{croName}\label{thm:cont}
When the grid \(\mathcal{H}^{t-1}\) becomes sufficiently dense, and points in \(\mathcal{H}^{t-1}\) are parameterized continuously, the summation in Thm.~\ref{thm:symm} converges to the corresponding integral,
\[
P_e^t 
\;=\; 
\frac{2}{L_{x}^{t-1}L_y^{t-1}} \int_{\mathcal{H}_1^t \setminus \mathcal{H}_2^t} 
Q\!\Bigl(\tfrac{\|\vh^t(\boldsymbol{x}) - \vh^t(\boldsymbol{x'})\|_2}{2\,\sigma}\Bigr) \, d\boldsymbol{x},
\]
where \(\boldsymbol{x'}\) denotes the symmetric “partner” of \(\boldsymbol{x}\) in the splitting dimension \(d^t\) and $\vh^t(\boldsymbol{x})$ is the continuous extension of $\vh^{t}_{\vk}$.   
\end{restatable}
%\xl{Do we need to mention somewhere how to generalize our algorithm to continuous search space? What likelihood do we calculate? Kaan: Can we still say we calculate infinitesimal likelihoods for infinitesimally seperated points numerically? I don't have a simulation for a truly continous search space} \textcolor{blue}{CF: we would need probability density of target at each continuous point given measurements and then integrate: lets not go into this and do after the deadline.}
See Appendix~\ref{appx:thm-cont} in \cite{Appendix} for a formal statement and proof.  Cor.~\ref{thm:cont} confirms that in the limit of very dense hypothesis grids, \(P_e^t\) under symmetric sensing can still be approximated by simple one-dimensional Gaussian tail integrals.

\subsection{Analysis in One-Dimensional Search Spaces}
\label{sec:1D-analysis}
Although our approach naturally extends to arbitrary dimensions, we find it interesting to adapt Algorithm~\ref{alg:iterative_binary_search} to a one-dimensional (1D) setting that allows easier visualization. 

\smallskip
\noindent \textbf{Setup.}
Let $\mathcal{H}^0 \subset \mathbb{R}$ be the evenly spaced grid of $n$ points in an interval of length $L$, specifically
\(
\mathcal{H}^0 
\,=\,
\Bigl\{\tfrac{(2i-1)\,L}{2n^{(0)}} : i=1,\dots,n^{(0)}\Bigr\}.
\)
At step $t$, the hypothesis set $\mathcal{H}^{t-1}$ is a subset of these grid points. We partition $\mathcal{H}^{t-1}$ into
$
\mathcal{H}_1^t 
=\{\,k : k \le m^t + \delta^t\Delta\}, 
$ and $
\mathcal{H}_2^t 
=\{\,k : k \ge m^t - \delta^t\Delta\},
$
where $m^t$ is the midpoint of $\mathcal{H}^{t-1}$, $\Delta=\frac{L^0}{n^{(0)}}$ is the spacing between points, and $\delta^t$ satisfies \eqref{eq:delta_identifier}.
Following the symmetric-measurement principle of Section~\ref{sec:symmetric}, sensors are activated at $\{\,s^t,\,L^{t-1}-s^t\}$, where $L^{t-1}$ is the interval length in the start of step $t$.

\smallskip
\begin{restatable}[{\em \em 1D Error Probability $P_e^t$ }]{corollary}{croName}  
\label{thm:1d}

\indent\emph{(a)~Discrete Grid.} 
At step $t$, let $|\mathcal{H}^{t-1}| = n^{(t-1)}$.  Then
\[
P_e^t 
\;=\;
\frac{2}{n^{(t-1)}} 
\sum_{\,k \in\, \mathcal{H}_1^t \setminus \mathcal{H}_2^t} 
Q\!\Bigl(\tfrac{d_{k}}{\sigma}\Bigr)
\;=\;
\frac{2}{n^{(t-1)}}\,
\sum_{\substack{k\le m^t - \delta^t\Delta}}
Q\!\Bigl(\tfrac{d_{k}}{\sigma}\Bigr),
\]
where $d_{k} = \tfrac{1}{2}\|\vh_{k}^t - \vh_{(L^{t-1} - k)}^t\|_2$ is half the distance (in measurement-space) between the hypothesis vectors associated with symmetric points $k$ and $L^{t-1} - k$.

\indent\emph{(b)~Continuous Limit.}
As $n\to\infty$, 
%the grid becomes arbitrarily fine and
\[
P_e^t 
\;=\; 
\frac{2}{L^{t-1}} \int_{\,0}^{\,L^{t-1}\left(\tfrac{1}{2}-\alpha\right)} 
Q\!\Bigl(\tfrac{\|\vh^t(x) - \vh^t(L^{t-1} - x)\|_2}{2\,\sigma}\Bigr)\,dx,
\]
where $L^{t}=\bigl(\tfrac12 + \alpha\bigr)L^{t-1}$.
\end{restatable}

\smallskip
\begin{boldremark}[Dependence on Amount of Overlap.]
\label{rmk:derivative}
When the sensor locations $\{s^t,\,L^{t-1} - s^t\}$ are fixed, one can examine how $P_e^t$ changes as $\alpha$ grows.  Differentiating $P_e^t$ yields
\[
\frac{\partial P_e^t}{\partial \alpha}
\;=\;
-\,{2}\,Q\Bigl(\tfrac{d\left(L^{t-1}\left(\tfrac{1}{2}-\alpha\right)\right)}{\sigma}\Bigr),
\]
where $d(x)=\|\vh(x) - \vh(L^{t-1}-x)\|_2 / 2$.  
Because $d(x)\ge0$, the maximum \emph{instantaneous} error reduction occurs at $\alpha=0$ and equals $-2\,Q(0)=-1$, after which further overlap provides diminishing returns. 
\end{boldremark}

\section{Evaluation}
\label{sec:evaluation}
We here present numerical evaluation results (delegating the detailed case specifications and additional plots in Appendix \ref{appx:plots} in \cite{Appendix}), and focusing on the 1D case. 
%We start by exploring the impact of overlap (parameter $\alpha$ on the tree depth and delay, then consider 

%In this section, we present numerical results for tree depth, sensor placement, and overall error probability as a function of tree depth as further analyses for the one-dimensional search space case in our algorithm. Tree depth results are provided for both continuous and discretized search spaces, while sensor placement and overall error probability results are focused on continuous search spaces.
\begin{figure}[t!]
\centering
\includegraphics[width=0.45\textwidth]{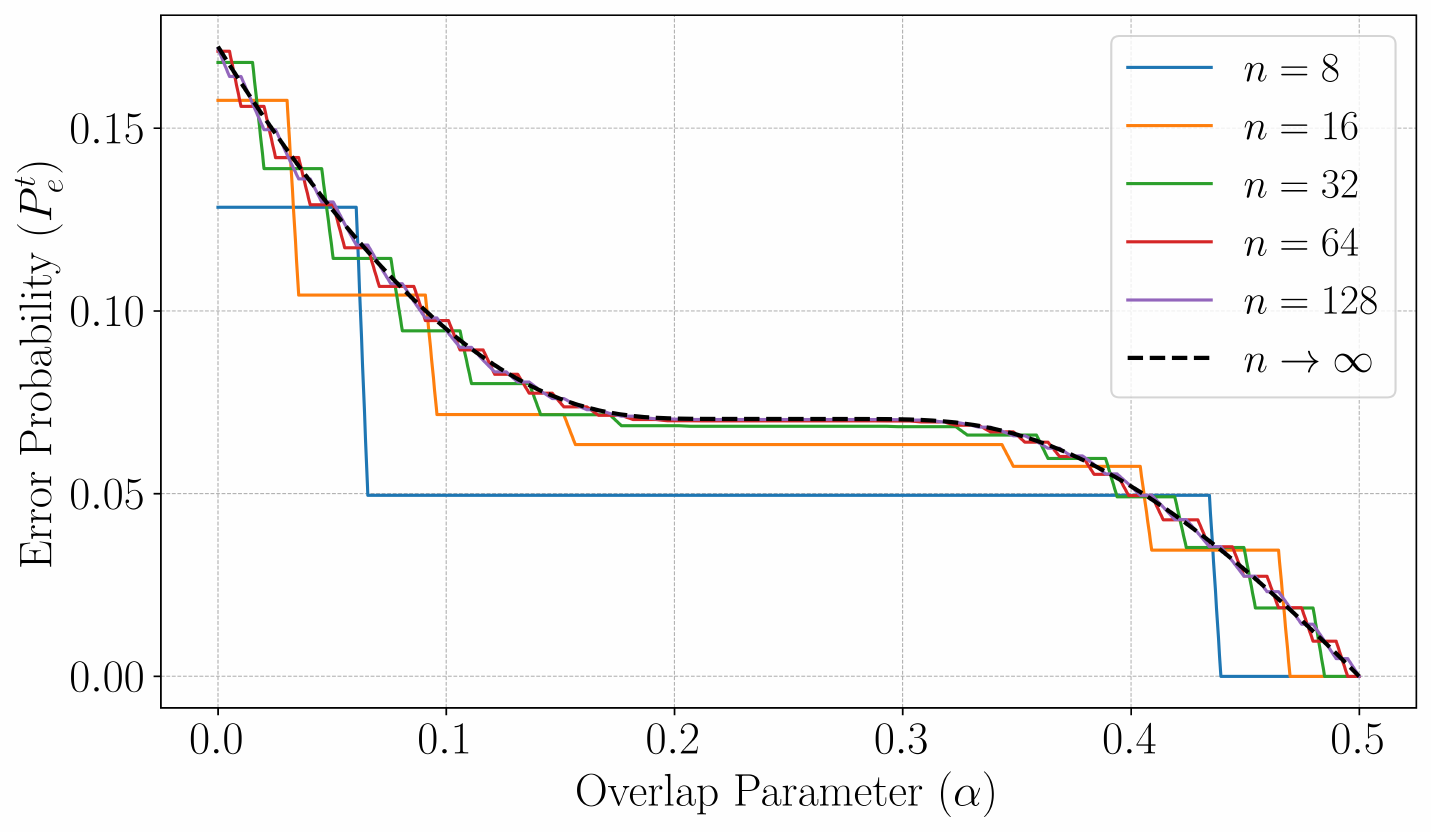}
\vspace{-.1in}
\caption{Error probability \(P_e^t\) vs.\ overlap parameter \(\alpha\) in a 1D setting, 
    for increasing \(n\) and fixed sensor location \(s^t=\tfrac{L^{t-1}}{4}\). 
    See Appendix~\ref{appx:plots}.\cite{Appendix}}
\vspace{-.1in}
\label{fig:error_vs_alpha}
\end{figure}

\begin{figure*}[t!]
    \centering
    % First Subfigure
    \subfloat[EA optimized and uniformly distributed sensor positions.]{%
        \includegraphics[width=0.45\textwidth]{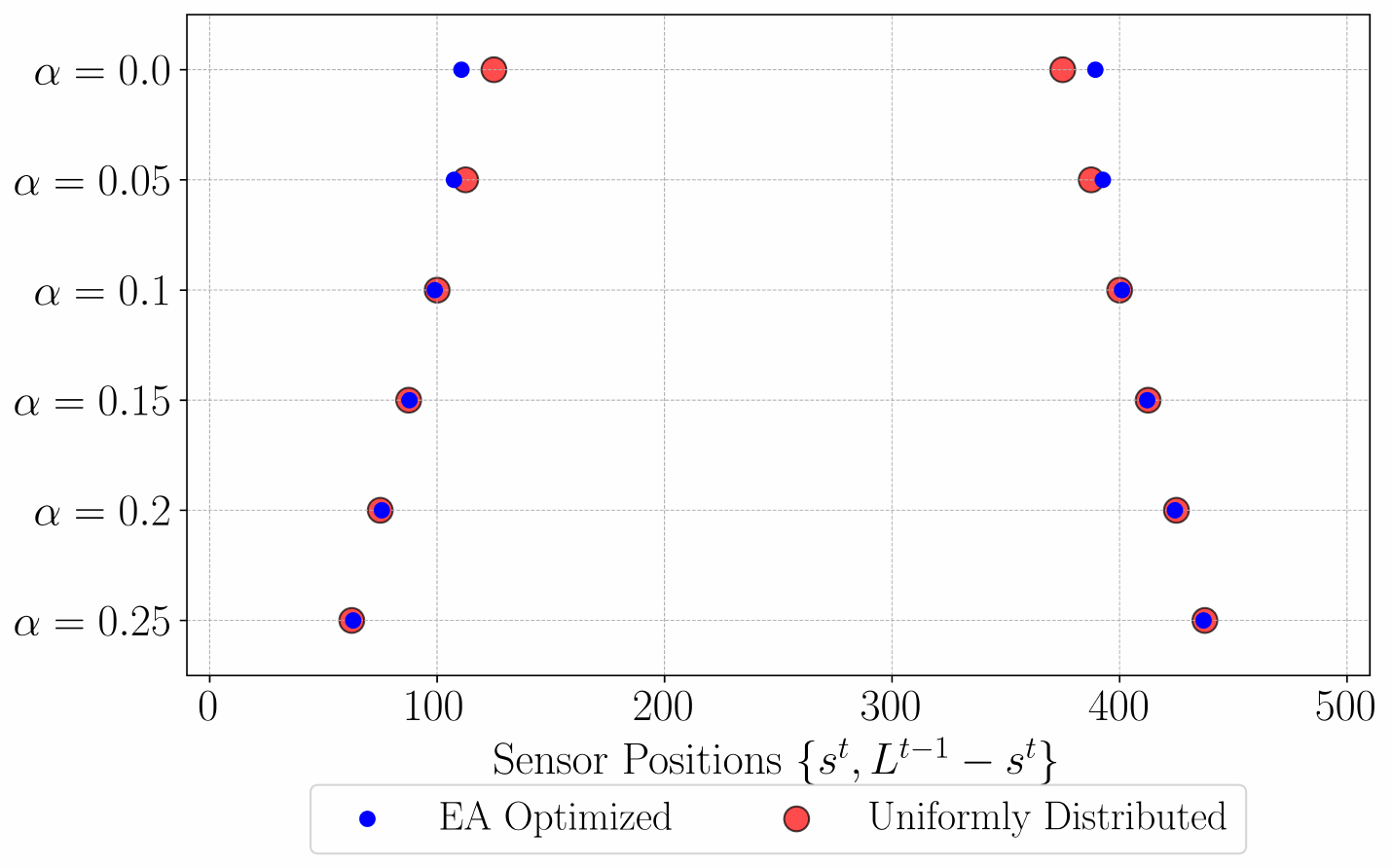}
        \label{fig:sensor_positions}
    }
    \hfil
    % Second Subfigure
    \subfloat[One step error probabilities $P_e^t$ for EA optimized and uniformly distributed sensor positions.]{%
        \includegraphics[width=0.45\textwidth]{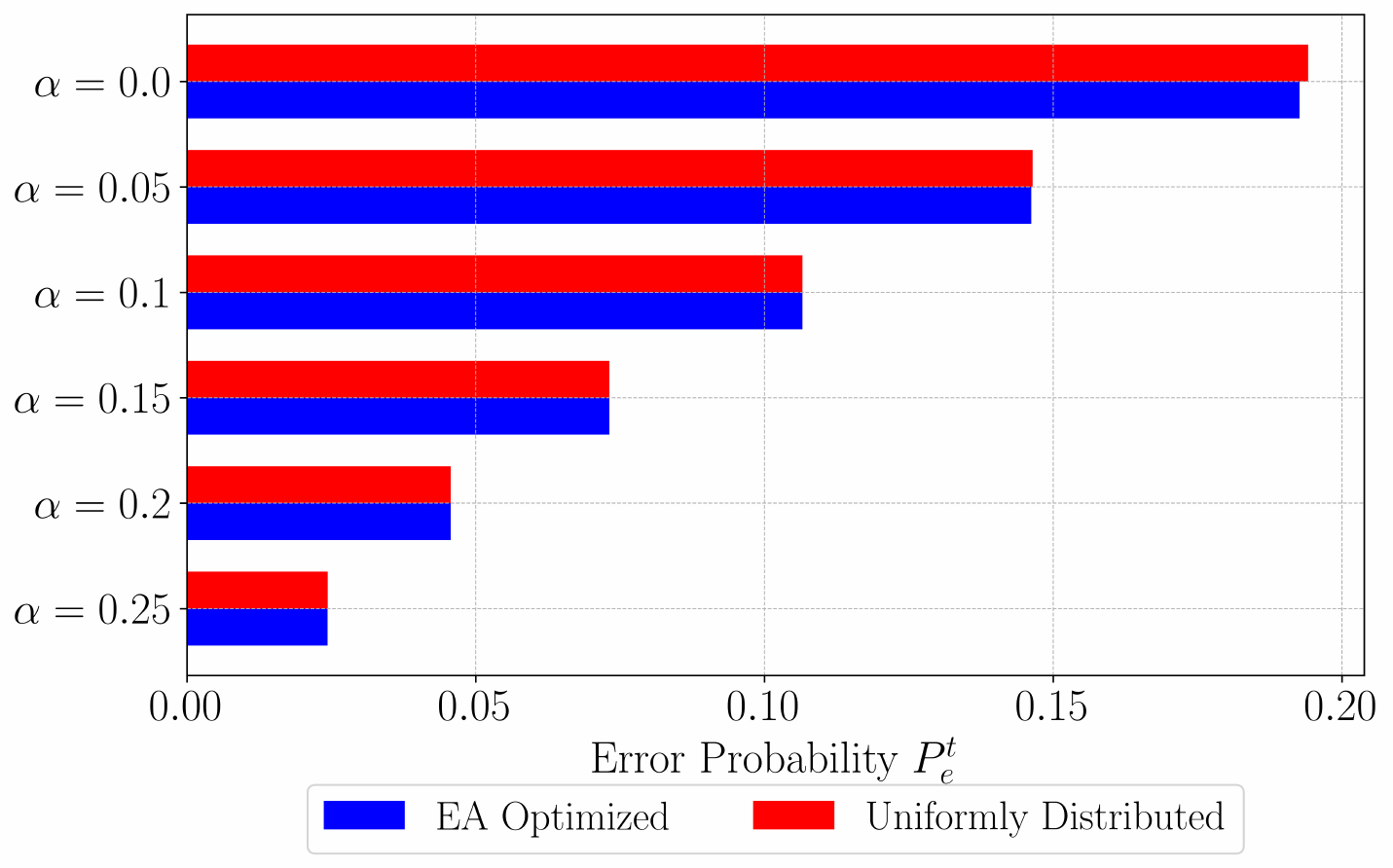}
        \label{fig:error_probabilities}
    }
    \caption{Comparison of sensor positions optimized with respect to $P_e^t$ with an evolutionary algorithm and with uniform distribution described in \ref{sec:num}}
    \vspace{-.1in}
    \label{fig:ea}
\end{figure*}

\medskip
\noindent{\em $\bullet$ Amount of Overlap vs Error Probability.}
Figure~\ref{fig:error_vs_alpha} depicts how $P_e^t$ varies with the overlap parameter $\alpha$ as $n$ increases.  All other parameters, including sensor placement, remain fixed (see Appendix~\ref{appx:plots} in \cite{Appendix} for details). The observed flattening for a middle range of the overlap parameters can be attributed to the specific configuration in sensor placement geometry and the propagation function $f(\cdot)$. Indeed, we show later in this section that a consistent decrease in error can be achieved when sensor placement is strategically adjusted as a function of the overlap parameter $\alpha$.
%for a fixed set of $s^t$.  
%We note that $s^t$ may also be optimized adaptively, as discussed in Section~\ref{sec:num}.
%\end{boldremark}

\medskip
\noindent{$\bullet$ \em Amount of Overlap vs. Tree Depth (Delay).}
Referring to (\ref{eq:delta_identifier}), it is possible to approximately find the dependence of the tree depth with respect to $\alpha$ in the discrete case by the approximation $n^{(0)}\left(\frac{1}{2} + \alpha\right)^\beta \approx n^{(\beta)}$ where $n^{(\beta)}$ refers to number of points left in the search space in the end of step $\beta$. Then we can approximate the tree depth $\beta$ as 
\begin{equation}
\beta \approx \operatorname{round}\left(\frac{\log n^{(\beta)} - \log{n^{(0)}}}{\log \left(\frac{1}{2} + \alpha\right)}\right).
\label{eq:approx_depth}
\end{equation}

\noindent The approximation  in a continuous search space becomes:
\begin{equation}
\beta = \left(\frac{\log L^{\beta} - \log{L}}{\log \left(\frac{1}{2} + \alpha\right)}\right). \nonumber
\end{equation}
Numerical evaluation over trees created with the partitioning rule in (\ref{eq:delta_identifier}) shows that (\ref{eq:approx_depth}) offers a good approximation of the tree depth 
(indicative plots can be found in Appendix \ref{appx:plots} in \cite{Appendix} for $n^{(\beta)} = 1$ and $n^{(0)} = 100,000$).
%shows that  of this approximated tree depth in (\ref{eq:approx_depth}) with a simulated iterative partitioning using the partitioning rule in (\ref{eq:delta_identifier}) applied to one dimensional search can be found in Appendix in \ref{appx:plots} \cite{Appendix} for $n^{(\beta)} = 1$ and $n = 100,000$. \textcolor{blue}{CF: Need to say what this comparison shows, for instance: We find that ****. Why are we plotting this?}\textcolor{red}{where we numerically show that (\ref{eq:approx_depth}) provides a good approximation for the tree depth of Algorithm \ref{alg:iterative_binary_search} even when the remaining search space on the output of the algorithm $\mathcal{H^\beta}$ is small. Kaan: To numerically show that the approximation is a good approximation, we can also remove the plot if unnecessary}

\begin{figure}[h!]
    \centering
    \includegraphics[width = 0.45 \textwidth]{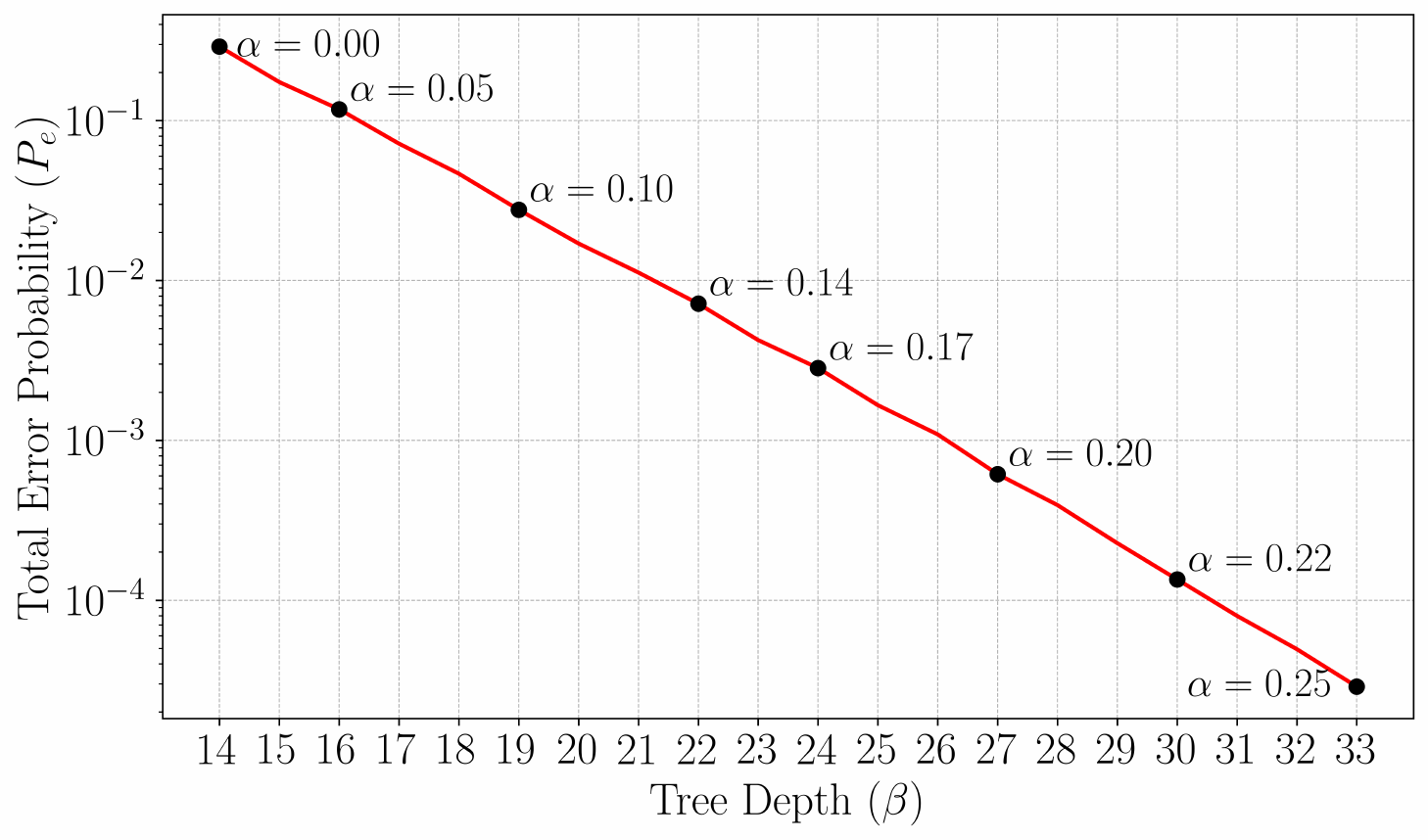}

    \caption{Error probability in the final output of the algorithm vs. tree depth $\beta$.}
    \label{fig:error_vs_beta}
\end{figure}

\medskip
\noindent{$\bullet$ \em Error Probability vs Tree Depth.}
Figure~\ref{fig:error_vs_beta} illustrate an  example of the trade off between the (end-to-end) error probability of the algorithm in (\ref{eq:tot_error}) and the decision tree depth assuming  uniformly distributed sensor positions with \( s^t = \frac{L^{t-1}}{2}\left(\frac{1}{2} - \alpha\right) \), (see Appendix \ref{appx:plots} \cite{Appendix} for simulation parameters). 
Consistent with our results for \( P_e^t \), we observe that even a small degree of overlap enables the algorithm to rapidly reduce the total error probability. This highlights a crucial trade-off: while the number of terms in (\ref{eq:tot_error}) increases with each iteration, i.e., we have more steps to make an error, this effect is mitigated by the significant reduction in \( P_e^t \) as the overlap increases and we are able to see a consistent logarithmic decline of the error probability with the tree depth.

\begin{figure}[h]
    \centering
    \includegraphics[width = 0.45\textwidth]{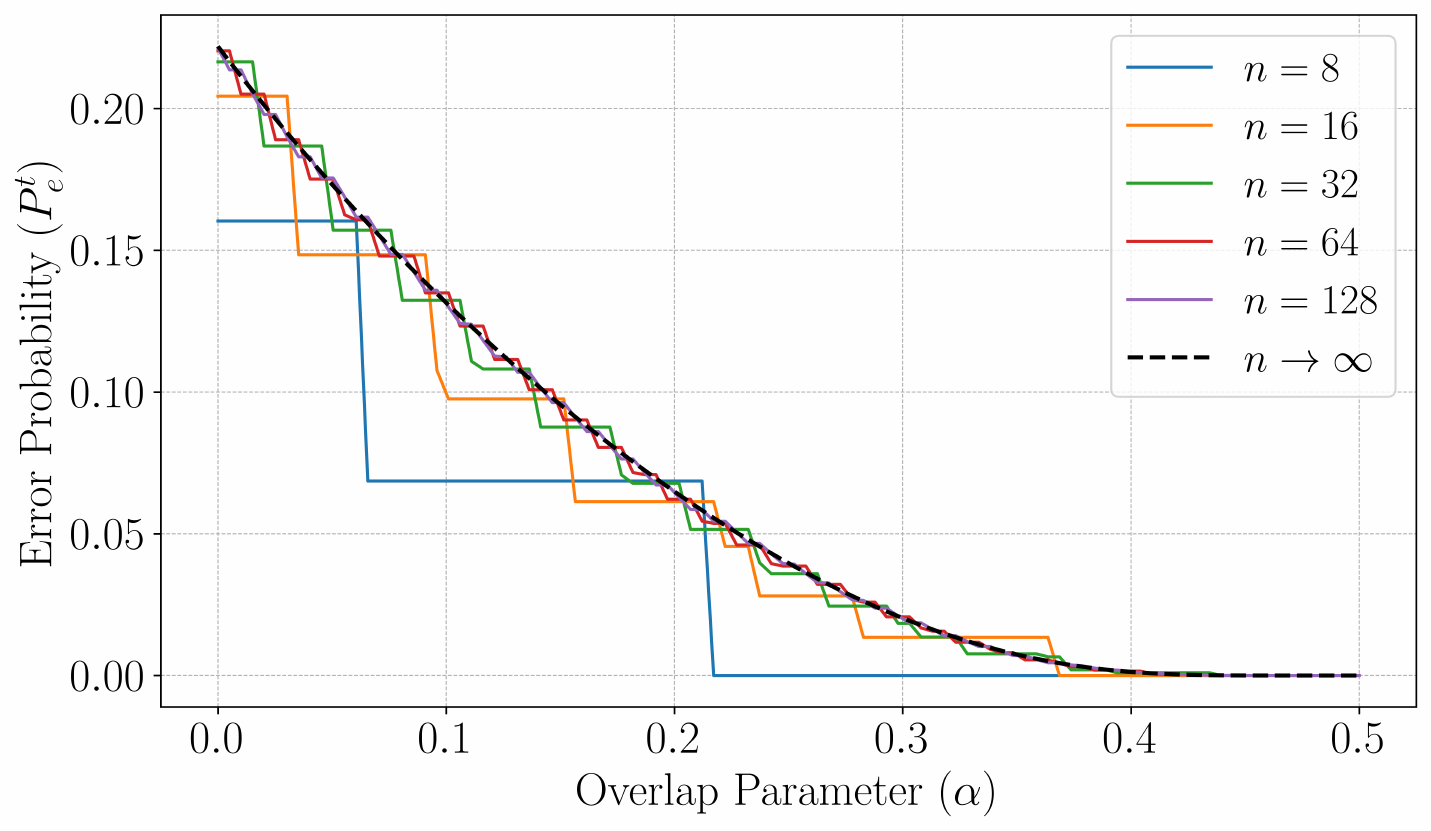}
    \caption{Error probability \(P_e^t\) vs.\ overlap parameter \(\alpha\) with sensor locations specified by \( s^t = \frac{L^{t-1}}{2}\left(\frac{1}{2} - \alpha\right) \).}
    \label{fig:error_vs_alpha_2}
\end{figure}

\noindent{\em $\bullet$  Sensor Placement Optimization.}
\label{sec:num} We here explore what would be an optimal sensor placement that would minimize our algorithm's error probability \( P_e^t \) at each iteration \( t \). 
%We note that works such as \cite{slv2014,flt2012} in sensor placement for localization typically employ techniques aimed at maximizing the Fisher information, thereby minimizing the mean squared error (MSE) of the estimation. In contrast, our objective focuses  optimizes a different criterion- the error probability \( P_e^t \) at each iteration \( t \). 
Because \( P_e^t \) is a non-smooth function of the sensor locations \( s^t \), we employ an Evolutionary Algorithm (EA) to find the sensor placements that minimize \( P_e^t \). 
The results, as shown in Fig.~\ref{fig:ea}, indicate that uniformly placing the sensors within the intervals \(\{[0, {L^{t-1}}\left(\frac{1}{2} - \alpha\right)],[{L^{t-1}}\left(\frac{1}{2} + \alpha\right),{L^{t-1}}]\}\), a.k.a. \( s^t = \frac{L^{t-1}}{2}\left(\frac{1}{2} - \alpha\right) \), serves as an effective approximation to the EA optimized placements. This observation supports the intuitive strategy of placing sensors near the midpoint of the exclusive regions of partitions to achieve balanced coverage. The accuracy of this approximation is particularly evident in scenarios with higher overlap, where the sensor placement closely matches the results obtained from EA.  
We note that applying this sensor placement rule 
%for the overlap parameter in question 
enables a consistent reduction in error probability. Figure~\ref{fig:error_vs_alpha_2} illustrates the change in the one step error probability $P_e^t$ when the fixed sensor placement rule in Figure~\ref{fig:error_vs_alpha} is updated to \( s^t = \frac{L^{t-1}}{2}\left(\frac{1}{2} - \alpha\right). \)

\vspace{0.15in}
\section{Conclusion}
\label{sec:conclusion}

We propose an overlapping binary search framework, where controlled intersections among partitions mitigate decision errors without losing the recursive structure of classical bisection, and where tuning the overlap parameter balances robustness and efficiency against noise. The approach applies to noisy sequential decision problems, illustrated by localization in sensor networks, with broader applications including medical diagnostics and crowdsourcing.
%For example, in adaptive disease testing, where each noisy test provides partial information about a patient's condition, a tunable search strategy would enable medical professionals to explicitly balance diagnostic speed against reliability, adapting to the clinical urgency or acceptable risk thresholds. 
Future work includes incorporating an adaptive number of measurements at each step, extending to high-dimensional spaces with computational constraints, and dynamically tuning $\alpha$ based on confidence or noise. Additional open questions include extending theoretical guarantees to asymmetric sensor placements and characterizing performance under more general noise models, including non-Gaussian, non-log-normal, and correlated settings.

\newpage
%\enlargethispage{-1.2cm}
\IEEEtriggeratref{14}
\bibliographystyle{IEEEtran}
\bibliography{reference}

\onecolumn % Switch to single-column layout
\appendices % Mark the start of the appendix section

\section{Proof of Lemma~\ref{thm:main}}
\label{appx:thm-main}

The Maximum Likelihood (ML) estimate of the target location \( \boldsymbol{l}_T \) at step \( t \) is:
\begin{equation*}
\boldsymbol{k}_{\mathrm{ML}}^t = \arg\max_{\boldsymbol{k} \in \mathcal{H}^{t-1}} \log p(\boldsymbol{z}^t \mid {\boldsymbol{l}_T} = \boldsymbol{k}),
\end{equation*}

where:

\begin{equation*}
p(\boldsymbol{z}^t \mid {\boldsymbol{l}_T} = \boldsymbol{k}) = \prod_{j=1}^{|\mathcal{S}^t|} \frac{1}{\sqrt{2\pi\sigma^2}} \exp\left(-\frac{(z_j^t - h_{\boldsymbol{k},j}^t)^2}{2\sigma^2}\right).
\end{equation*}

The log-likelihood is:
\begin{equation*}
\ell(\boldsymbol{k}) = -\frac{1}{2\sigma^2} \sum_{j=1}^{|\mathcal{S}^t|} \left(z_j^t - h_{\boldsymbol{k},j}^t\right)^2 - \frac{|S^t|}{2}\log 2\pi\sigma^2.
\end{equation*}

Thus, maximizing \( \ell(\boldsymbol{k}) \) is equivalent to finding the point $\vk \in \mathcal{H}^{t-1}$ such that:
\begin{equation*}
\boldsymbol{k}_{\mathrm{ML}}^t = \arg\min_{\boldsymbol{k} \in \mathcal{H}^t} \|\boldsymbol{z}^t - \boldsymbol{h}_{\boldsymbol{k}}^t\|_2^2.
\end{equation*}

Thus, the ML estimate of $\boldsymbol{l}_T$ corresponds to selecting \( \boldsymbol{h}_{\boldsymbol{k}}^t \) closest to \( \boldsymbol{z}^t \) in Euclidean distance.
Therefore we consider the Voronoi cells associated with each \( \hyp \) in \( \mathcal{H}^{t-1} \). The Voronoi cell for hypothesis \( \vh_{\vk}^t \) is defined as:
\begin{equation*}
V_{\vk}^t = \{\boldsymbol{x} \mid \|\boldsymbol{x} - \vh_{\vk}^t\|_2 \leq \|\boldsymbol{x} - \vh_{\vk'}^t\|_2, \, \forall \vk' \neq \vk; \vk' \in \mathcal{H}^{t-1} \}.
\nonumber
\end{equation*}

Each region also can be expressed as a union of half spaces:
\begin{equation*}
    V_{\vk}^t = \{\boldsymbol{x} \mid \boldsymbol{x}^\top (\vh_{\vk}^t - \vh_{\vk'}^t) \geq \frac{1}{2} \left( \|\vh_{\vk}^t\|_2^2 - \|\vh_{\vk'}^t\|_2^2 \right), \forall \vk' \neq \vk; \vk' \in \mathcal{H}^{t-1} \}.
\end{equation*}

According to the decision rule in Algorithm \ref{alg:iterative_binary_search}, the error probability at step $t$ is
\begin{equation*}
    P_e^t = \Pr(\vk_\mathrm{ML} \notin \mathcal{H}_m^t) \;=\; \Pr \left( \bigcup_{\vk \in \mathcal{H}_m^t} \left\{ \boldsymbol{z}^t \notin V_{\vk}^t\right\}\right) = \Pr\!\biggl(\,\boldsymbol{z}^t \,\notin\, 
\bigcup_{\vk \in \mathcal{H}_m^t} V_{\vk}^t\biggr)
\end{equation*}

if $\boldsymbol{l_T} \notin \mathcal{H}^t_1 \cap \mathcal{H}^t_2$, where $m \in \{1,2\}$ such that $\boldsymbol{l}_T \in H^t_m$.

{\hfill$\square$\\}

\section{Proof of Theorem~\ref{thm:symm}}
\label{appx:thm-symm}

The symmetric placement of sensors \(\mathcal{S}^t = \{\boldsymbol{s}_1^t, \boldsymbol{s}_2^t\}\) around $m_{d^t}$ can be formally defined as:
\begin{equation*}
\boldsymbol{s}_2^t = 
\begin{cases}
(u, L_y^t - v), \quad \text{if } \mathrm{d}^t = y, \\
(L_x^t - u, v), \quad \text{if } \mathrm{d}^t = x,
\end{cases}
\end{equation*}

given $\boldsymbol{s}_1^t = (u,v)$. 
% with $u \in [0,\frac{1}{2}L_x^{t-1}] $ and $ v \in [0,\frac{1}{2}L_y^{t-1}]$.

To prove Thm. \ref{thm:symm}, we consider pairs \( (\vk,\vk') \) in the search space \( \mathcal{H}^{t-1} \) whose symmetry is imposed similarly to the sensor placement rule. The symmetric pair of $\vk=(u,v)$ is given by:
\begin{equation}
\vk' = 
\begin{cases}
(u, L_y^t - v), \quad \text{if } \mathrm{d}^t = y, \\
(L_x^t - u, v), \quad \text{if } \mathrm{d}^t = x,
\end{cases}
\label{eq:symm-pairs}
\end{equation}

where \( (u, v) \in \mathcal{H}^{t-1} \). We also note that there exists a unique $\vk'$ for every $\vk \in \mathcal{H}^{t-1}$, which directly follows from the definition of $\mathcal{H}^{0}$ and the partitioning rule in \ref{eq:subregion}. 
The hypothesis vectors for symmetric points satisfy:
\begin{equation*}
h_{\vk, 1} = h_{\vk', 2}, \quad h_{\vk, 2} = h_{\vk', 1}.
\end{equation*}

\begin{figure}
    \centering
    \includegraphics[width = 0.4\textwidth]{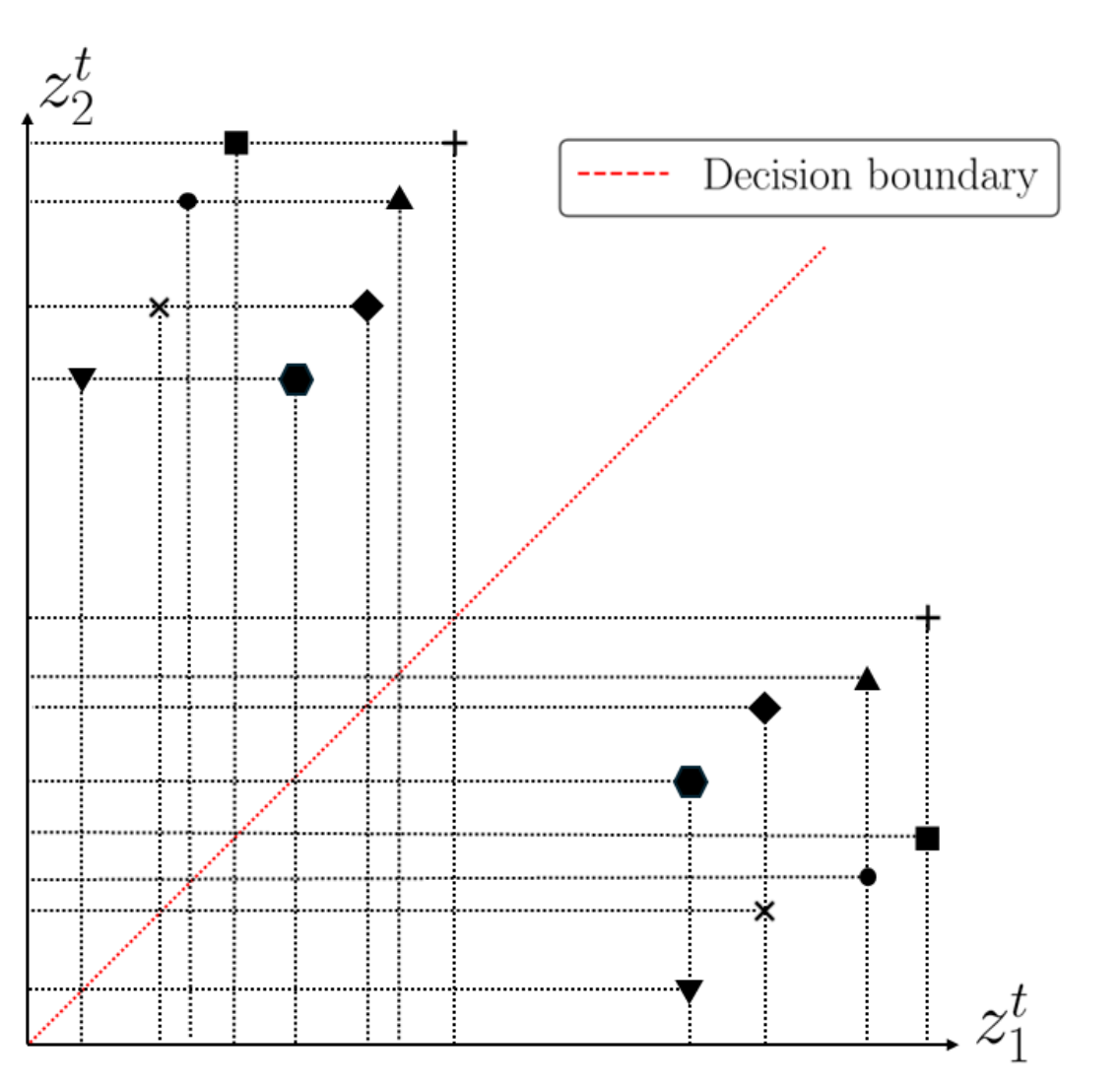}
    \caption{Example $(\hyp, \vh_{\vk'}^t)$ pairs shown in $(|S_t|=2)$-dimensional space for an hypothesis set of $|\mathcal{H}^{t-1}|=16$}
    \label{fig:hk-pairs}
\end{figure}

Because of this symmetric property, the decision boundary separating the halves \( \mathcal{H}_1^t \) and \( \mathcal{H}_2^t \) converges into a single hyperplane that is the symmetry axis $\boldsymbol{w}^\top \boldsymbol{z}^t = b$ where: 

\begin{equation*}
\boldsymbol{w} = \frac{1}{\sqrt{2}}[1,-1]^\top, \quad
b = 0.
\end{equation*}
An example of such setup is demonstrated in Figure \ref{fig:hk-pairs}.

If $\boldsymbol{l_T = \vk} $ for $\exists \vk \in \mathcal{H}^{t-1}$, then the measurements are generated with $\boldsymbol{z}^t = \vh_{\vk}^t + \boldsymbol{\eta}^t$ where \( \boldsymbol{\eta}^t \sim \mathcal{N}(0, \sigma^2 \mathbf{I}) \). The distance of $\vh_{\vk}^t$ to the decision boundary can be computed efficiently by utilizing the hypothesis vectors $\vh_{\vk}$ and $ \vh_{\vk'}$ associated with the symmetric pair $(\vk,\vk')$ as: 

\begin{equation*}
d_{\vk} = \frac{\|\vh_{\vk}^t - \vh_{\vk'}^t\|_2}{2}.
\end{equation*}

% Let $\boldsymbol{w}_\perp = \frac{1}{\sqrt{2}}[1,1]^\top$ be the perpendicular direction to the decision boundary. 
The probability of error at step $t$ is the probability that the noise projection to the normal vector of the decision boundary $\eta_{\perp} = \boldsymbol{w}^\top \boldsymbol{\eta}^t \sim \mathcal{N}(0, \sigma^2)$ is higher in magnitude than $d_{\vk}$ and is in the direction towards the decision boundary:

\begin{equation*}
\Pr(\boldsymbol{l}_T \notin \mathcal{H}^{t} \mid \boldsymbol{l}_T = \vk) = \Pr\left(\eta_{\perp} \leq -d_{\vk}\right) = \Pr\left(\eta_{\perp} \geq d_{\vk}\right) = Q\left(\frac{d_{\vk}}{\sigma}\right)
\end{equation*}

where the error event is either $\left\{\eta_{\perp} \leq -d_{\vk}\right\} $ or $ \left\{\eta_{\perp} \geq d_{\vk}\right\}$ depending on the location of $\vh_k^t$ in $|\mathcal{S}^t|$-dimensional space.

\medskip
With a uniform prior distribution for \( \boldsymbol{l}_T \) over \( \mathcal{H}^{t-1} \), the error probability at step \( t \) over all possible $\vk \in \mathcal{H}^{t-1}$ is given by:
\begin{align*}
P_e^t &= \Pr\{\boldsymbol{l}_T \notin \mathcal{H}^t \mid \boldsymbol{l}_T \in \mathcal{H}^{t-1}\} = \sum_{\vk \in \mathcal{H}^{t-1}} 
\Pr(\boldsymbol{l}_T = \vk)
\Pr(\boldsymbol{l}_T \notin \mathcal{H}^t \mid \boldsymbol{l}_T = \vk)  \\
&= \frac{1}{|\mathcal{H}^{t-1}|}\Bigl(\sum_{\vk \in \mathcal{H}_1^t \setminus \mathcal{H}^t_2} \Pr(\boldsymbol{l}_T \notin \mathcal{H}^t \mid \boldsymbol{l}_T = \vk) + \sum_{\vk \in \mathcal{H}_2^t \setminus \mathcal{H}^t_1} \Pr(\boldsymbol{l}_T \notin \mathcal{H}^t \mid \boldsymbol{l}_T = \vk) \Bigr) \nonumber\\
&= \frac{2}{|\mathcal{H}^{t-1}|} \sum_{\vk \in \mathcal{H}_1^t \setminus \mathcal{H}^t_2} \Pr(\boldsymbol{l}_T \notin \mathcal{H}^t \mid \boldsymbol{l}_T = \vk) = \frac{2}{|\mathcal{H}^{t-1}|} 
\sum_{\vk \in \mathcal{H}_1^t \setminus \mathcal{H}_2^t} 
Q\left(\frac{d_{\vk}}{\sigma}\right) \\
&= \frac{2}{n_x^{(t-1)} n_y^{(t-1)}} 
\sum_{\vk \in \mathcal{H}_1^t \setminus \mathcal{H}_2^t} 
Q\left(\frac{d_{\vk}}{\sigma}\right),
\end{align*}
where \( \Pr(\boldsymbol{l}_T \notin \mathcal{H}^t \mid \boldsymbol{l}_T = \vk) = 0 \) for all \( \vk \in \mathcal{H}_1^t \cap \mathcal{H}_2^t \). 
{\hfill$\square$\\}

\newpage
\section{Proof of Corollary~\ref{thm:cont}}
\label{appx:thm-cont}

We prove the case where \(d^t = y\) for the sake of simplicity in notation; the case where \(d^t = x\) follows directly.

For large enough $\{n^{t-1}_x,n_y^{t-1}\}$, the decision rule in Equation \ref{eq:delta_identifier}, converges to $n_{d^t}^{(t)} = \left(\frac{1}{2} + \alpha\right)n^{(t-1)}_{d^t}$. Thus for large $\{n^{t-1}_x,n_y^{t-1}\}$,
\[
P_e^t 
\;=\; 
\frac{2}{n_x^{(t-1)}\,n_y^{(t-1)}} 
\sum_{\vk \,\in\, \mathcal{H}_1^t \setminus \mathcal{H}_2^t} 
Q\!\Bigl(\tfrac{d_{\vk}}{\sigma}\Bigr) = \frac{2}{n_x^{(t-1)} n_y^{(t-1)}} \sum_{k_x=0}^{n_x^{(t-1)}} \; \sum_{k_y=0}^{\left(\frac{1}{2} - \alpha\right)n_y^{(t-1)}} Q\left(\frac{d_{(k_x, k_y)}}{\sigma}\right),
\]

We here define the continuous extension of the utilized hypothesis vectors $\vh_{\vk}^t$ as $\vh^t(\boldsymbol{x})$ whose $i$-the element is $h^t(\boldsymbol{x})_{i} = f\left(\|\boldsymbol{s}_i^t - \boldsymbol{x}\|_2\right)$ defined over the continuous region of $[0,L_x^{t-1}] \times [0,L_y^{t-1}]$. Subsequently $d(\boldsymbol{x}) = \frac{1}{2}\|\vh^t(\boldsymbol{x}) - \vh^t(\boldsymbol{x'})\|_2$. where $\boldsymbol{x'}$ is the symmetric pair of $\boldsymbol{x}$ dictated by (\ref{eq:symm-pairs}).

\medskip

The summation can be rewritten as:
\begin{equation}
    P_e^t = \frac{2}{L_x^{t-1} L_y^{t-1}} \sum_{k_x=0}^{n_x^{(t-1)}} \sum_{k_y=0}^{\left(\frac{1}{2} - \alpha\right)n_y^{(t-1)}} Q\left(\frac{d(\vk)}{\sigma}\right) \Delta k_x \Delta k_y,
\label{eq:rects}
\end{equation}

where \(\Delta k_x = \frac{L_x^{t-1}}{n_x^{(t-1)}}\) and \(\Delta k_y = \frac{L_y^{t-1}}{n_y^{(t-1)}}\) and $\vk = [k_x,k_y]^T$. 

\medskip

Under mild regularity assumptions on the propagation function \( f(\cdot) \), \( d(\vk) \) is continuous, bounded, and integrable. Consequently, as \( n_x^{(t-1)}, n_y^{(t-1)} \to \infty \), (\ref{eq:rects}) converges to the corresponding Riemann integral:
\[
\lim_{n_x^{(t-1)},n_y^{(t-1)}\rightarrow \infty} P_e^t = \frac{2}{L_x^{t-1} L_y^{t-1}}  \int_0^{L_x^{t-1}} \int_0^{\left(\frac{1}{2} - \alpha\right)L_y^{t-1}} Q\left(\frac{d(\vk)}{\sigma}\right) \,dk_y\,dk_x =
\frac{2}{L_x^{t-1} L_y^{t-1}} \int_{\mathcal{H}^t_1 \setminus \mathcal{H}_2^t} Q\left(\frac{d(\boldsymbol{x})}{\sigma}\right) d \boldsymbol{x}.
\]

{\hfill$\square$\\}

\section{Parameters Specification and Extended Results}
\label{appx:plots}
\subsection{Amount of Overlap vs Error Probability}

This subsection outlines the simulation parameters and methodology used to generate the plot in Figure~\ref{fig:error_vs_alpha}. For notational simplicity in this evaluation, we define the number of points in the remaining search space \( \mathcal{H}^{t-1} \) at step $t$ as \( n = n^{(t-1)} \) and compute the corresponding $P_e^t$. The key simulation parameters are summarized in Table~\ref{tab:sim_params-1} with the utilized propagation function $f(\cdot)$ given by:

\begin{equation}
10\log_{10}f(d) = P_0 - \eta \cdot 10 \cdot \log_{10}(d + \epsilon),
\label{eq:prop_function}
\end{equation}
where \(d\) is the distance from the target location to a sensor, and \(\epsilon\) regularizes the logarithmic term. 

\begin{table}[h!]
\centering
\caption{Parameters utilized in generating Figure \ref{fig:error_vs_alpha}.}
\renewcommand{\arraystretch}{1.3}
\begin{tabular}{|c|c|l|}
\hline
\textbf{Parameter}       & \textbf{Value(s)}               & \textbf{Description}                           \\ \hline
\(L^{t-1}\)                    & 500                            & Length of the search space.                   \\ \hline
\(n\)                    & 8, 16, 32, 64, 128             & Number of remaining hypotheses at step $t$.    \\ \hline
\(P_0\)                  & 20                             & Reference signal strength in dB.                \\ \hline
\(\eta\)                 & 1                              & Path loss exponent.                           \\ \hline
\(\sigma\)               & 1/$\sqrt{2}$                              & Standard deviation of measurement noise.       \\ \hline
\(\epsilon\)             & 10                             & Regularization parameter. \\ \hline
\(\alpha\)               & \([0, 0.5]\)                  & Overlap parameter, uniformly sampled.         \\ \hline
$S^t$ (digitized)         & \(\lfloor n/4 \rfloor\ , \lfloor 3n/4 \rfloor\)       & Placement of sensors in the search space.     \\ \hline
$S^t$ (continuous)           & \(L/4, 3L/4\)       & Placement of sensors when $n \rightarrow \infty$  \\ \hline
\end{tabular}
\vspace{0.3cm}

\label{tab:sim_params-1}
\end{table}

We also note that, although we initially define the overlap parameter $\alpha$ in the range $[0,\frac{1}{4})$ to ensure convergence to any $n^{(\beta)}$ including $n^{(\beta)} = 1$ with partitioning rule (\ref{eq:delta_identifier}), Thm. \ref{thm:symm} holds for $\alpha \in [0,0.5]$ which is the range used for Figures \ref{fig:error_vs_alpha},\ref{fig:error_vs_alpha_2}.

\subsection{Amount of Overlap vs. Tree Depth (Delay)}
\begin{figure}[htbp]
\centering
\includegraphics[width=0.45\textwidth]{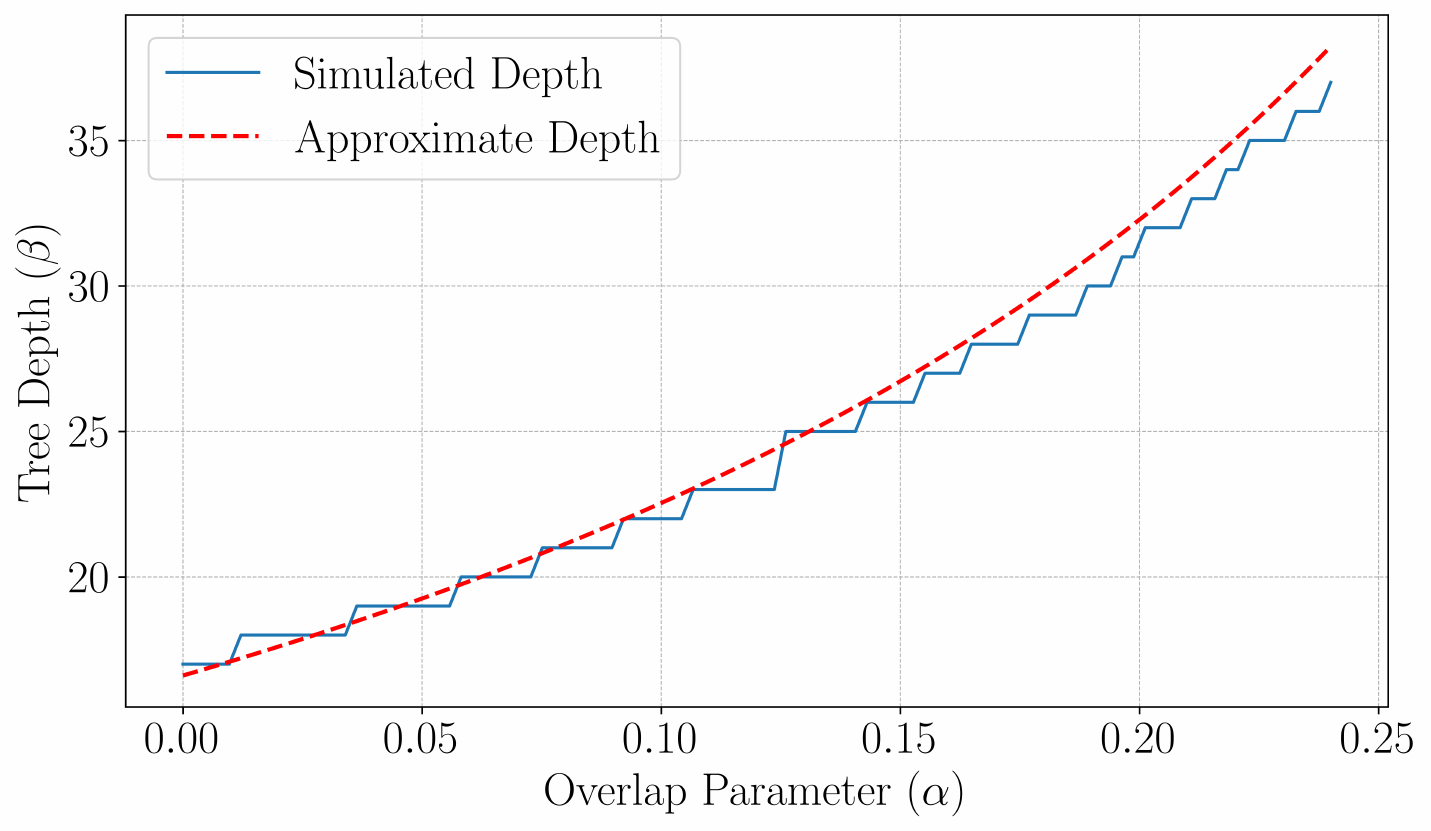}
\caption{Tree depth parameter \(\beta\) vs.\ overlap parameter \(\alpha\).}
\label{fig:tree_depth_vs_alpha}
\end{figure}

In this subsection, we present the numerical validation and approximation for the relationship between the overlap parameter \(\alpha\) and the tree depth \(\beta\), as described in the main paper. The plot in Fig.~\ref{fig:tree_depth_vs_alpha} compares the simulated tree depth (numerical) and the approximate formula derived for \(\beta\) in (\ref{eq:approx_depth}).

The simulation iteratively calculates the depth \(\beta\) by starting with \(n^{(0)} = 100,000\) hypotheses and applying the partitioning rule in (\ref{eq:delta_identifier}).

\subsection{Error Probability vs. Tree Depth}
\label{sec:error_vs_beta}

This subsection outlines the parameters used to generate the plot in Figure~\ref{fig:error_vs_beta}. The overlap parameter \(\alpha\) in this evaluation is dynamically adjusted for each desired tree depth $\beta$.

The numerical computation utilizes the parameters summarized in Table~\ref{tab:sim_params-2}, where the utilized propagation function is given in (\ref{eq:prop_function}).

\begin{table}[htbp]
\centering
\renewcommand{\arraystretch}{1.3}
\caption{Parameters Utilized in Generating Figure \ref{fig:error_vs_beta}}

\begin{tabular}{|c|c|l|}
\hline
\textbf{Parameter}       & \textbf{Value(s)}                       & \textbf{Description}                           \\ \hline
\(P_0\)                  & \(20\)                              & Reference signal strength in dB.                \\ \hline
\(\eta\)                 & \(1\)                                  & Path loss exponent.                           \\ \hline
\(\sigma\)               & \(0.2\)                                & Standard deviation of measurement noise.       \\ \hline
\(\epsilon\)             & \(10\)                                 & Regularization parameter. \\ \hline
\(n^{(0)}\)                  & \(2^{14}\)                             & Initial size of the hypothesis set.           \\ \hline
\(n^{(\beta)}\)                  & \(2^{7}\)                             & Final size of the hypothesis set.           \\ \hline
\(\beta\)                & \([14, 33]\) & Range of tree depths. \\ \hline

\(L^0\)                    & \(500\)                                & Length of the initial search space.                   \\ \hline
$S^t$          & \(\left\{\lfloor(1/2-\alpha) \cdot n^{(t-1)}\rfloor, \lfloor(3/2+\alpha) \cdot n^{(t-1)}\rfloor \right\}\) & Sensor positions dynamically adjusted with \(\alpha\). \\ \hline
\end{tabular}
\vspace{0.3cm}

\label{tab:sim_params-2}
\end{table}
\newpage
\subsection{Sensor Placement Optimization}

This subsection details the simulation parameters used to approximate the optimal sensor placement locations minimizing the one-step error probability \( P_e^t \) in continuous space according to Corollary \ref{thm:1d}, as presented in Figure~\ref{fig:ea}. The optimization employs an Evolutionary Algorithm (EA), with results compared against a uniformly spaced heuristic placement. The parameters used in this evaluation are summarized in Table~\ref{tab:sim_params_sensor_placement} with the utilized propagation function given in (\ref{eq:prop_function}).
\begin{table}[htbp]
\centering
\renewcommand{\arraystretch}{1.3}
\caption{Parameters Utilized in Generating Figure \ref{fig:ea}.}
\begin{tabular}{|c|c|l|}
\hline
\textbf{Parameter}       & \textbf{Value(s)}                       & \textbf{Description}                           \\ \hline
\(L^{t-1}\)             & \(500\)                                & Length of the search space.                   \\ \hline
\(P_0\)                & \(20\)                                & Reference signal strength in dB.              \\ \hline
\(\eta\)               & \(1\)                                  & Path loss exponent.                           \\ \hline
\(\sigma\)             & \(1\)                                  & Standard deviation of measurement noise.      \\ \hline
\(\epsilon\)           & \(10^{-4}\)                            & Regularization parameter.                     \\ \hline
\multicolumn{3}{|c|}{\textbf{Evolutionary Algorithm (EA) Parameters}} \\ \hline
Population size         & \(50\)                                 & Number of individuals in the EA.              \\ \hline
Generations            & \(50\)                                 & Number of iterations for the EA optimization. \\ \hline
Mutation probability   & \(0.2\)                                & Mutation probability in the EA.               \\ \hline
Mutation method        & Gaussian                               & Mutation method applied.                      \\ \hline
Mutation mean (\(\mu\)) & \(0\)                                & Mean value for Gaussian mutation.             \\ \hline
Mutation std (\(\sigma\)) & \(10\)                      & Standard deviation for Gaussian mutation.     \\ \hline
Mutation probability per gene  & \(0.2\) & Probability of mutating an individual gene.  \\ \hline
Crossover probability  & \(0.5\)                                & Probability of crossover in the EA.           \\ \hline
Crossover method       & Blend                                  & Crossover method applied.                     \\ \hline
Crossover blend alpha  & \(0.5\)                                & Blend ratio for crossover.                    \\ \hline
Selection method       & Tournament selection                   & Selection strategy.                           \\ \hline
Tournament size        & \(3\)                                  & Number of individuals in each tournament.     \\ \hline
\end{tabular}
\vspace{0.3cm}
\label{tab:sim_params_sensor_placement}
\end{table}

\end{document}